\documentclass[10pt,letterpaper,twocolumn]{article} 

\usepackage{ol2}
\usepackage[draft]{hyperref}
\usepackage{amsmath}

\begin{document}

\twocolumn[ 

\title{Non-Bloch $\mathcal{PT}$ symmetry breaking in non-Hermitian Photonic Quantum Walks}


\author{Stefano Longhi}

\address{Dipartimento di Fisica, Politecnico di Milano and Istituto di Fotonica e Nanotecnologie del Consiglio Nazionale delle Ricerche, Piazza L. da Vinci 32, I-20133 Milano, Italy (stefano.longhi@polimi.it)}

\begin{abstract}
 A hallmark of topological band theory in periodic media is that bulk properties are not affected by boundary conditions. Remarkably, in certain non-Hermitian lattices the bulk properties are largely affected by boundaries, leading to such major effects as the non-Hermitian skin effect and violation of the bulk-boundary correspondence. Here we unveil that non-unitary discrete-time quantum walks of photons in systems involving gain and loss show rather generally non-Bloch parity-time ($\mathcal{PT}$) symmetry breaking phase transitions, and suggest a bulk probing method to detect such boundary-driven phase transitions. 
 \end{abstract}

 ] 

{\it Introduction.} A hallmark of topological band theory in periodic media is that bulk properties, such as Bloch energy bands and gaps, are not affected by boundary conditions, and Bloch band invariants can predict edge effects via the bulk-boundary correspondence \cite{r1}. However, several recent studies showed that in certain non-Hermitian systems the bulk properties are largely affected by boundaries, leading to such major effects as the non-Hermitian skin effect (NHSE), i.e. the squeezing of bulk modes to the edges, anomalous edge states, the violation of the bulk-boundary correspondence based on Bloch topological invariants, and symmetry-breaking phase transitions of non-Bloch bands \cite{r2,r3,r4,r5,r6,r7,r8,r9,r10,r11,r12,r13,r14,r15,r15b,r16,r18,r17,r17bis}.  Such discoveries challenge the current wisdom of topological order and their explanation requires suitable generalization of Bloch band theory and topological invariants \cite{r4,r5,r8,r9,r12,r13}. The bulk bands in systems with  open boundary conditions (OBC) can considerably differ from those of systems with periodic
boundary conditions (PBC). While the latter are defined by ordinary Bloch band theory, the former are non-Bloch bands that require the Bloch quasi-momentum to become complex and to vary on a generalized Brillouin zone \cite{r4,r5,r8,r9,r13,r14,r15}. A major consequence is that distinct symmetry breaking phase transitions can be observed when considering Bloch and non-Bloch bands \cite{r2,r4,r7,r14,r15}, i.e. {\it bulk} symmetries and symmetry breaking phase transitions in non-Hermitian systems can be affected by {\it boundaries}.\\
Photonics provides an experimentally accessible platform to test non-Hermitian topological systems (see e.g. \cite{r17,r19,r20,r21,r22,r23,r24,r25,r25b,r26,r27} and references therein), with potential applications to novel laser design \cite{r28,r29,r30}, quantum light \cite{r31} and light steering \cite{r32}, to mention a few. Discrete-time quantum walks (QWs) of photons, realized in different optical settings and probed with ether classical or quantum light \cite{r17bis,r17,r23,r24,r33,r34,r35,r36,r37,r37b,r37c}, are versatile systems to investigate topological order in Floquet dynamics and to implement non-Hermitian effects. While Bloch band $\mathcal{PT}$ symmetry breaking phase transitions via exceptional points have been observed in photonic QW several years ago \cite{r33},  only very recently the ability of QWs to show non-Bloch topological features has been suggested, with the experimental demonstration of the NHSE and bulk-boundary correspondence breakdown in a QW of single photons \cite{r17}. However, the occurrence of $\mathcal{PT}$ symmetry breaking phase transitions of non-Bloch bands and their experimental signatures remain largely unexplored.\\
In this Letter we unveil that, besides NHSE, non-Hermitian photonic QWs show rather generally non-Bloch $\mathcal{PT}$ symmetry-breaking phase transitions, which can be probed in the bulk from Lyapunov exponent measurements of the QW dynamics. \par
  \begin{figure}[htbp]
\centerline{\includegraphics[width=8.6cm]{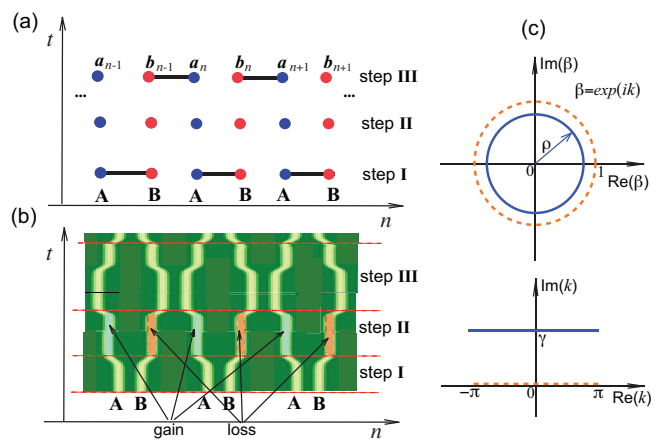}} \caption{ \small
(Color online) (a) Schematic of the three-step non-Hermitian quantum walk on a one-dimensional lattice. (b) Physical implementation in waveguide arrays. (c) Representation of the ordinary (dashed curves) and generalized (solid curves) Brillouin zones in $\beta=\exp(ik)$ plane (upper panel) and Bloch wave number $k$ plane (lower panel). In $\beta$ plane, the ordinary Brillouin zone is the unit circle $C_{\beta}$, while the extended Brillouin zone $\tilde{C}_{\beta}$ is the circle of radius $\rho=\exp(-\gamma)$, where $\gamma$ is the gain/loss parameter introduced in step II. The QW becomes unitary, with ordinary and generalized Brillouin zone conciding, for $\gamma=0$.}
\end{figure} 
{\it Non-Hermitian photonic quantum walks: model and $\mathcal{PT}$ symmetry.} We consider the three-step discrete-time photonic QW  on a one-dimensional lattice schematically shown in Fig.1(a). The QW can be implemented using integrated optical directional couplers, in which two sublattices A and B of a waveguide array with a slowly-curved optical axis are alternately coupled over one modulation cycle [Fig.1(b)]; see e.g. \cite{r35}. In steps I and III, optical power is exchanged between pairs of closely-spaced waveguides in the two sublattices, whereas in the intermediate step II all waveguides are widely-spaced and decoupled, while alternating gain (optical amplification) and loss (optical attenuation) is introduced in the two sublattices. We note that the non-unitary QW  can be also implemented using different optical settings, such as bulk polarizer beam splitters and quarter-wave plates \cite{r17,r23} or fiber network loops \cite{r37b} --the states A and B being the polarization states of the photons-- or chiral light in which the lattice sites are associated with photonics states carrying different quanta of orbital angular momentum and the states A and B are mapped onto spin angular momentum \cite{r34}. Let us indicate by $a_n(t)$ and $b_n(t)$ the field amplitudes in the two sublattice waveguides at propagation distance $t$.
The evolution of the amplitudes over one modulation cycle (of period $T=1$) is governed by the map
\begin{eqnarray}
a_n(t+1) & = & \alpha \exp(\gamma) a_n(t)+\beta \exp(\gamma) b_n(t) \\
& + & \delta \exp(-\gamma) a_{n-1}(t)+\sigma \exp(-\gamma)b_{n-1}(t) \nonumber \\
b_n(t+1) & = &  \alpha \exp(-\gamma) b_n(t)+\beta \exp(-\gamma) a_n(t) \\
& + & \delta \exp(\gamma) b_{n+1}(t)+\sigma \exp(\gamma)a_{n+1}(t) \nonumber 
\end{eqnarray}
where $\gamma>0$ is the balanced gain/loss parameter introduced in step II, $\alpha \equiv \cos \theta_1 \cos \theta_2$, $\beta \equiv i \sin \theta_1 \cos \theta_2$, $\delta \equiv - \sin \theta_1 \sin \theta_2$, $\sigma \equiv i \cos \theta_1 \sin \theta_2$, and $\theta_1$, $\theta_2$ are the rotation angles of the coin state (i.e. effective waveguide couplings) in steps I and III, respectively. 
For PBC, owing to the discrete translational invariance one can write $(a_n(t),b_n(t))^T=(A(t),B(t))^T \exp(ikn)$, where $k$ is the Bloch wave number which varies in the first Brillouin zone $-\pi \leq k < \pi$, i.e. $\beta \equiv \exp(ik)$ varies on the unit circle $C_{\beta}$ of complex $\beta$ plane ($|\beta|=1$). The change of amplitudes $A(t)$, $B(t)$ over one modulation cycle reads $(A(1),B(1))^T=U(k) (A(0),B(0))^T$, where the Floquet propagator $U(k)$ in momentum space can be written as
\begin{eqnarray}
U(k) & = & S^{-1}(k)R(\theta_2)S(k) L R(\theta_1)=\exp(-ik \sigma_3/2) \nonumber \\
& \times & \exp(i \theta_2 \sigma_1) \exp[(ik/2+\gamma) \sigma_3] \exp(i \theta_1 \sigma_1)
\end{eqnarray} 
and $\sigma_{1,2,3}$ are the Pauli matrices.
In the above equation, $S(k)$ is the shift operator in momentum space, given by
\begin{equation}
S(k) =\left(
\begin{array}{cc}
\exp(ik/2) & 0 \\
0 & \exp(-ik/2)
\end{array}
\right)=\exp(i k \sigma_3 /2),
\end{equation}
$R(\theta)$ is the rotation matrix for an angle $\theta$,
\begin{equation}
R(\theta)= \left(
\begin{array}{cc}
\cos \theta & i \sin \theta \\
i \sin \theta & \cos \theta
\end{array}
\right)= \exp(i \theta \sigma_1),
\end{equation}
$\theta_1$, $\theta_2$ are the rotation angles in steps I and III, respectively, and
\begin{equation}
L= \left(
\begin{array}{cc}
\exp(\gamma) & 0 \\
0 & \exp(-\gamma)
\end{array}
\right)=\exp(\gamma \sigma_3)
\end{equation}
describes balanced light amplification and attenuation in sublattices A and B at step II. The elements of the propagator $\mathcal{U}$ read explicitly
\begin{eqnarray}
\mathcal{U}_{11}(k) & = & \exp(\gamma) \cos \theta_1 \cos \theta_2-\exp(-\gamma-ik) \sin \theta_1 \sin \theta_2 \nonumber \\
\mathcal{U}_{12}(k) & = & i\exp(\gamma) \sin \theta_1 \cos \theta_2+i \exp(-\gamma-ik) \cos \theta_1 \sin \theta_2 \nonumber \\
\mathcal{U}_{21}(k) & = & i\exp(-\gamma) \sin \theta_1 \cos \theta_2+i \exp(\gamma+ik) \cos \theta_1 \sin \theta_2 \;\;\;\; \nonumber \\
\mathcal{U}_{22}(k) & = & \exp(-\gamma) \cos \theta_1 \cos \theta_2-\exp(\gamma+ik) \sin \theta_1 \sin \theta_2 \nonumber .
\end{eqnarray}
An effective non-Hermitian Hamiltonian in momentum space for PBC can be introduced as usual through the relation $U(k)=\exp(-iH_{eff})$. The QW dynamics under $\mathcal{U}$ can be regarded as a stroboscopic map of the non-unitary dynamics governed by $H_{eff}$. It should be noted that the form of $\mathcal{U}(k)$ (and thus of $H_{eff}$) is not unique since it depends on the chosen basis of states A and B. In particular, for a rotation $R(\varphi)$ of the A/B basis by an angle $\varphi$, the Floquet propagator $U(k)$ is transformed into $U^{\prime}(k)=R(\varphi)U(k)R^{-1}(\varphi)$. As the quasi energies are clearly unchanged, the rotation is useful to unveil hidden symmetries in the non-unitary quantum walk \cite{r38}. In particular, the Floquet propagator $U^{\prime}(k)$ possesses $\mathcal{PT}$ symmetry (under a suitable definition of parity $\mathcal{P}$ and time reversal $\mathcal{T}$ operators) provided that \cite{r38}
\begin{equation}
\mathcal{PT}U^{\prime}(k)=(U^{\prime}(k))^{ -1} \mathcal{PT},
\end{equation}   
i.e. $\mathcal{PT} H_{eff}^{\prime}(k)=H_{eff}^{\prime}(k) \mathcal{PT}$.
\\
\par
{\it Non-Hermitian skin effect and non-Bloch $\mathcal{PT}$ symmetry breaking phase transitions.} While the Floquet propagator $U(k)$ is non-unitary for $\gamma \neq 0$, one has ${\rm{det}}U(k)=1$. Using such a property,
the quasi energies $\epsilon_{1,2}(k)$ of the system with PBC, i.e. the eigenvalues of $H_{eff}$, can be readily calculated and read
\begin{equation}
\epsilon_{1,2}(k)= \pm \phi(k)
\end{equation}
where the complex angle $\phi(k)$ is defined via the relation
\begin{eqnarray}
\cos  \phi(k)  & \equiv &  \frac{U_{11}(k)+U_{22}(k)}{2} \\
& = & \cos \theta_1 \cos \theta_2 \cosh \gamma- \sin \theta_1 \sin \theta_2 \cosh (\gamma+ik). \nonumber
\end{eqnarray}
 \begin{figure}[htbp]
\centerline{\includegraphics[width=8.4cm]{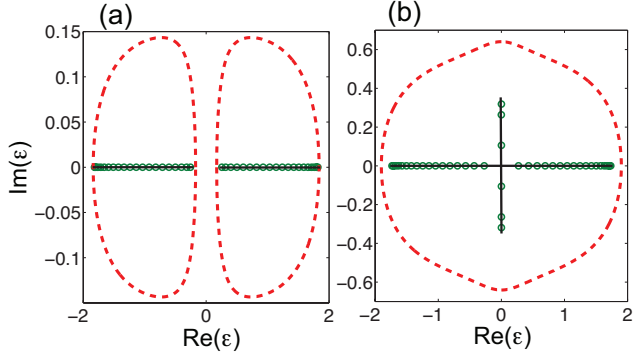}} \caption{ \small
(Color online) Quasi-energy spectrum of the non-Hermitian QW for parameter values $\theta_1=\pi /3$, $\theta_2=\pi /4$ and for (a) $\gamma=0.2$ (unbroken $\mathcal{PT}$ symmetric phase of non-Bloch bands), and (b) $\gamma=0.7$ (broken $\mathcal{PT}$ symmetric phase of non-Bloch bands). Dashed curves, describing closed loops, are the quasi energies of the QW with PBC [Eqs.(8) and (9) with $k$ real], whereas solid curves, describing open arcs, are the quasi energies for the QW with OBC [Eq.(10) with $q$ real]. Solid circles are the numerically-computed quasi energies of the Floquet propagator $U_{2N}$ in real space for a QW comprising $N=20$ dimers with OBC. The non-Bloch $\mathcal{PT}$ symmetry breaking phase transition occurs at $\gamma=\gamma_{th}=0.4356$ [Eq.(11)].}
\end{figure} 
and $k$ varies in the ordinary Brilloiun zone [dashed curve in Fig.1(c)].
When the system comprises $N$ dimers with OBC, the quasi energy spectrum should be numerically computed from the $2N \times 2N$ Floquet propagator $\mathcal{U}_{2N}$ in real space with appropriate boundary conditions. Interestingly, contrary to similar non-unitary QWs considered in previous works \cite{r33,r38}, in our QW the quasi energy spectrum for OBC differs from the one of PBC because of the NSHE \cite{r17}; examples of PBC and OBC quasi energy spectra are shown in Fig.2.
This result can be readily proven using the saddle-point criterion \cite{r15}, looking at the position in complex $\beta$ plane of the saddle points of quasi energies $\epsilon_{1,2}(\beta)$: if the saddle points $\beta_s$ of $\epsilon_{1,2}(\beta)$ do not lie on the unit circle $C_{\beta}$, then the OBC and PBC quasi energies are distinct and the bulk eigenstates are squeezed at the edges for OBC (NSHE). From Eqs.(8) and (9) one finds two saddle points at $\beta_s= \pm \exp(-\gamma)$ with modulus less than one, indicating the existence of the NSHE.\\ 
The bulk OBC quasi energy spectrum is again given by Eq.(8) and (9), but the Bloch wave number $k$ becomes complex and $\beta \equiv \exp(ik)$ varies on the generalized Brillouin zone $\tilde{C}_{\beta}$ \cite{r4,r5,r8,r9,r13}. The generalized Brillouin zone is determined with the procedure detailed in previous works \cite{r4,r5,r8,r9}, and the saddle points $\beta_s$ belong to $\tilde{C}_{\beta}$ \cite{r15}. From the form of Eq.(9), it readily follows that $\tilde{C}_{\beta}$ is the circle of radius $\rho=\exp(-\gamma)$, i.e. $k=i \gamma+q$ with $q$ real and $-\pi \leq q < \pi$ [Fig.1(c), solid curve]. In fact, the conditions of generalized Brillouin zone $\epsilon_{1,2}(\beta_1)=\epsilon_{1,2}(\beta_2)$ with $| \beta_1|=|\beta_2|$ \cite{r4,r5,r8,r9} are satisfied by letting $\beta_1=i \gamma+q$ and $\beta_2=i \gamma-q$ with $q$ real. Therefore, the quasi energy spectrum for OBC reads
\begin{eqnarray}
\epsilon_{1,2}(q) & = &  \pm \phi(q+i\gamma) \\
& = & {\rm {acos}} \left\{  \cos \theta_1 \cos \theta_2 \cosh \gamma- \sin \theta_1 \sin \theta_2 \cos q \right\}. \nonumber
\end{eqnarray}
Interestingly, while the quasi energy spectrum for PBC is always complex [Eqs.(8) and (9) and Fig.2] for any non vanishing value of the non-Hermitian parameter $\gamma$, the quasi energy spectrum for OBC [Eq.(10) and Fig.2] shows a  phase transition, from an entirely real quasi energy spectrum to complex conjugate quasi energies, as $\gamma$ increases above the threshold value $\gamma_{th}$ defined by the relation
\begin{equation}
\cosh \gamma_{th}=   \frac{ 1-|\sin \theta_1 \sin \theta_2|}{|\cos \theta_1 \cos \theta_2|}.
\end{equation}
 \begin{figure}[htbp]
\centerline{\includegraphics[width=8.4cm]{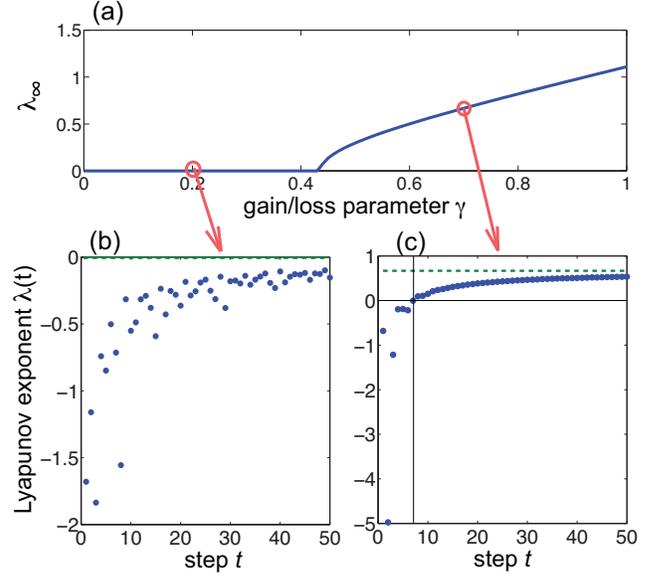}} \caption{ \small
(Color online) (a) Behavior of the asymptotic value of Lyapunov exponent $\lambda_{\infty}$ versus gain/loss parameter $\gamma$ in a QW for parameter values $\theta_1=\pi/3$ and $\theta_2=\pi/4$. Symmetry breaking phase transition, corresponding to an increase of $\lambda_{\infty}$ above zero, is observed for $\gamma> \gamma_{th}$, where $\gamma_{th}=0.4356$ is given by Eq.(11). Panels (b) and (c) show the nuemrically-computed Lyapunov exponent $\lambda(t)$ versus step number $t$ [Eq.(12)] for (b) $\gamma=0.2$, and (c) $\gamma=0.7$. The dashed horizontal lines define the asymptotic values $\lambda_{\infty}$.} 
\end{figure} 
Equation (11) is readily obtained after letting the argument of the ${\rm acos}$ function on the right hand side of Eq.(10) equal to one (in modulus) at the center ($q=0$) or at the edge ($q= \pm \pi$) of the Brillouin zone.
Note that, for $|\theta_{1}| \neq |\theta_2|$, $\gamma_{th}$ is strictly positive, with $\gamma_{th} \rightarrow 0$ as $\theta_{2} \rightarrow \pm \theta_1$ and $\gamma_{th} \rightarrow \infty$ as either $\theta_1$or $\theta_2$ gets close to $ \pm \pi/2$.\\ 
The transition from an entirely real to complex quasi energy spectrum can be viewed as a $\mathcal{PT}$ symmetry breaking phase transition of the Floquet propagator $U^{\prime}(k)$ in momentum space {\em on the generalized Brillouin zone} ($k=q+i \gamma$, $q$ real). In fact, it can be readily shown that rotation by the angle $\varphi= \theta/2+\pi/4$ of the state basis transforms the Floquet propagator $U(k)$ [Eq.(3)] into $U^{\prime}(k)=R(\varphi)U(k)R^{-1}(\varphi)$ satisfying Eq.(7) when $k$ varies on the generalized Brillouin zone and with parity and time-reversal operators defined by $\mathcal{P}=\sigma_1$ and $\mathcal{T}=\sigma_1 \mathcal{K}$ ($\mathcal{K}$ is the elementwise complex conjugation operation). Note that, contrary to other non unitary QWs \cite{r38}, in our model $U^{\prime}(k)$ does not possess $\mathcal{PT}$ symmetry when $k$ varies on the ordinary Brillouin zone, and indeed in the QW with PBC one does not observe any symmetry breaking phase transition of Bloch bands (Fig.2). Therefore, $\mathcal{PT}$ symmetry breaking observed in a QW with OBC is a {\em non-Bloch} phase transition.  Note also that, while the non-Hermitian QW defined by the non unitary Floquet propagator (3) always shows the NHSE \cite{r17}, non-Bloch phase transitions arise solely for an {\em unbalanced} QW , i.e. for $|\theta_1| \neq |\theta_2|$.\\
 \par
{\it Probing non-Bloch phase transitions.}  In an experimental setting, the appearance of non-Bloch phase transitions can be detected looking at the asymptotic QW dynamics in the bulk (i.e. far from edges), measured by the Lyapunov exponent of the map defined by Eqs.(1) and (2) \cite{r15}. At initial step $t=0$, let us excite the photonic structure with a photon in site $n=0$ of sublattice A, i.e. $a_n(0)=\delta_{n,0}$ and $b_n(0)=0$, and let us measure as successive steps $t=1,2,3,...$ the growth rate
\begin{equation}
\lambda(t)=\frac{1}{t} \log \left| a_0(t)\right|^2.
\end{equation}  
Following a procedure similar  to the one presented in \cite{r15}, it can be shown that, for large $t$, $\lambda(t)$ converges to the asymptotic value $\lambda_{\infty}$ given by
\begin{equation}
\lambda_{\infty}= \left\{
\begin{array}{l}
0  \;\;\;\; \;\;\;\; \;\;\;\; (\gamma<\gamma_{th} )\\
{2 \; \rm acosh} \left\{ | \cos \theta_1 \cos \theta_2 | \cosh \gamma +|\sin \theta_1 \sin \theta_2 | \right\}  \\
 \;\;\;\; \;\;\;\;\;\;\;\;\;\;\; (\gamma>\gamma_{th} )
\end{array}
\right.
\end{equation}
Therefore the non-Bloch $\mathcal{PT}$ symmetry breaking phase transition is signaled by the increase of $\lambda_{\infty}$ above zero [Fig.3(a)].  Examples of the numerically-computed behaviors of $\lambda(t)$ versus step number $t$, for $\gamma$ either below or above the symmetry breaking threshold $\gamma_{th}$, are shown in Figs.3(b) and (c). Note that for $\gamma$ sufficiently above $\gamma_{th}$, as in  Fig.3(c),
the positiveness of $\lambda(t)$, and thus the signature of the symmetry breaking, is observed after less than $ \sim 10$ periods (which is within experimental accessibility), while to reach the asymptotic value $\lambda_{\infty}$ much more steps would be required.\\
As a final comment, it should be mentioned that, while in our model we assumed balanced gain and loss in sublattices A and B, similar results hold for an entirely passive system, in which for example in step II sublattice B is lossy with loss parameter $\gamma$ while sublattice A is transparent (no loss, no gain). This situation, which could be experimentally more accessible since amplification is not required, would correspond to a down-shift of the quasi energy spectra by $\gamma/2$ along the imaginary axis of energy, while the non-Bloch phase transition would correspond to the so-called passive $\mathcal{PT}$ symmetry breaking \cite{r39}.\\ 
\par
{\it Conclusions}
Non-unitary quantum walks  of photons have been recently shown to provide a beautiful and experimentally accessible platform to investigate topological order in non-Hermitian Floquet systems, with the observation of bulk-boundary correspondence failure and squeezing of bulk eigenstates to the edges (NHSE) \cite{r17}. Here we have shown that non-unitary photonic quantum walks can provide a very promising tool  for the observation of another intriguing phenomenon in non-Hermitian lattices, i.e. symmetry breaking phase transitions of non-Bloch bands. Non-Bloch symmetry breaking phase transitions are bulk phase transitions driven by the boundary conditions. They are observed when the QW occurs on a linear chain with OBC, which forces the Bloch wave number to span an extended Brillouin zone, while they are not observed in a system with PBC, where the Bloch wave number spans the ordinary Brillouin zone. This is because the non-unitary Floquet propagator possesses certain symmetries when considered on the generalized Brillouin zone, but not on the ordinary Brillouin zone. The present results are expected to stimulate further theoretical and experimental investigations on the rapidly growing and fascinating area of non-Hermitian topological phases, with potential interest beyond photonics

\newpage
 {\bf References with full titles}\\
 \\
 \noindent
1. A. Bansil, H. Lin, and T. Das, {\it Colloquium: Topological band theory}, Rev. Mod. Phys. {\bf 88}, 021004 (2016).\\
2. T. E. Lee, {\it Anomalous Edge State in a Non-Hermitian Lattice}, Phys. Rev. Lett. {\bf 116}, 133903 (2016).\\
3. Y. Xiong, {\it Why Does Bulk Boundary Correspondence Fail in Some Non-Hermitian Topological Models}, J. Phys. Commun. {\bf 2}, 035043 (2018).\\
4. S. Yao and Z. Wang, {\it Edge States and Topological Invariants of Non-Hermitian Systems}, Phys. Rev. Lett. {\bf 121}, 086803 (2018).\\
5. S. Yao, F. Song, and Z.Wang, {\it Non-Hermitian Chern Bands}, Phys. Rev. Lett. {\bf 121}, 136802 (2018).\\
6.  F.K. Kunst, E. Edvardsson, J.C. Budich, and E.J. Bergholtz, {\it Biorthogonal Bulk-Boundary Correspondence in Non-Hermitian Systems}, Phys. Rev. Lett. {\bf 121}, 026808 (2018).\\
7. V. M. Martinez Alvarez, J. E. Barrios Vargas, and L. E. F. Foa Torres, {\it Non-Hermitian robust edge states in one dimension: Anomalous localization and eigenspace condensation 
at exceptional points}, Phys. Rev. B {\bf 97}, 121401 (2018).\\
8. C.H. Lee and R. Thomale, {\it Anatomy of skin modes and topology in non-Hermitian systems}, Phys. Rev. B {\bf 99}, 201103 (2019).\\
9. K. Yokomizo and S. Murakami, {\it Bloch Band Theory for Non-Hermitian Systems}, Phys. Rev. Lett. {\bf 123}, 066404 (2019).\\
10. C.H. Lee, L. Li, and J. Gong, {\it Hybrid higher-order skin-topological modes in non-reciprocal systems}, Phys. Rev. Lett. {\bf 123},  016805 (2019).\\
11. T.S. Deng and W. Yi, {\it Non-Bloch topological invariants in a non-Hermitian domain-wall system}, Phys. Rev. B {\bf 100}, 035102 (2019).\\
12. A. Ghatak and T. Das, {\it New topological invariants in non-Hermitian systems}, J. Phys.: Condens. Matter {\bf 31}, 263001 (2019).\\
13. F. Song, S. Yao, and Z. Wang, {\it Non-Hermitian Topological Invariants in Real Space}, arXiv:1905.02211v1 (2019).\\
14. T. Liu, Y.-R. Zhang, Q. Ai, Z. Gong, K. Kawabata, M. Ueda, and F. Nori, {\it Second-order topological phases in non-Hermitian systems}, Phys.
Rev. Lett. {\bf 122}, 076801 (2019).\\
15. L. Jin and Z. Song, {\it Bulk-boundary correspondence in a non-Hermitian system in one dimension with chiral inversion symmetry},
Phys. Rev. B {\bf 99}, 081103 (2019).\\
16. S. Longhi, {\it Probing non-Hermitian skin effect and non-Bloch phase transitions}, Phys. Rev. Res. {\bf 1}, 023013 (2019).\\
17. T. Hofmann, T. Helbig, F. Schindler, N. Salgo, M. Brzezinska, M. Greiter, T. Kiessling, D. Wolf, A. Vollhardt, A. Kabasi, C. H. Lee, A. Bilusic, R. Thomale, and T. Neupert, 
{\it Reciprocal skin effect and its realization in a topolectrical circuit}, arXiv:1908.02759v1 (2019).\\
18. T. Helbig, T. Hofmann, S. Imhof, M. Abdelghany, T. Kiessling, L.W. Molenkamp, C. H. Lee, A. Szameit, M. Greiter, and R. Thomale, {\it Observation of bulk boundary correspondence breakdown
in topolectrical circuits}, arXiv:1907.11562v1 (2019).\\
19. B. Wang, T. Chen, and X. Zhang, {\it Observation of novel robust edge states without bulkboundary
correspondence in non-Hermitian quantum walks}, arXiv 1906.06676 (2019).\\
20. T. Deng, K. Wang, G. Zhu, Z. Wang, W. Yi, and P. Xue,
{\it Observation of non-Hermitian bulk-boundary correspondence in quantum dynamics}, arXiv:1907.12566v1 (2019).\\
21. T. Ozawa, H.M. Price, A. Amo, N. Goldman, M. Hafezi, L. Lu, M. Rechtsman, D. Schuster, J. Simon, O. Zilberberg, and I. Carusotto, {\it Topological photonics}, Rev. Mod. Phys. {\bf 91}, 015006  (2019).\\
22. J. M. Zeuner, M. C. Rechtsman, Y. Plotnik, Y. Lumer, S. Nolte, M. S. Rudner, M. Segev, and A. Szameit, {\it Observation of a topological transition in the bulk of a non-Hermitian system}, Phys. Rev. Lett. {\bf 115}, 040402 (2015).\\
23. C. Poli, M. Bellec, U. Kuhl, F. Mortessagne, and H. Schomerus, {\it Selective enhancement of topologically induced interface states in a dielectric resonator chain}, Nature Commun. {\bf 6}, 6710 (2015).\\
24. S. Weimann, M. Kremer, Y. Plotnik, Y. Lumer, S. Nolte, K. G. Makris, M. Segev, M.C. Rechtsman, and A. Szameit, {\it Topologically protected bound states in photonic parity-time-symmetric crystals}, Nat. Mater. {\bf 16}, 433?438 (2017).\\
25. L. Xiao, X. Zhan, Z. H. Bian, K. K. Wang, X. Zhang, X. P. Wang, J. Li, K. Mochizuki, D. Kim, N. Kawakami, W. Yi, H. Obuse, B. C. Sanders, and P. Xue, {\it Observation of topological edge states in
parity-time-symmetric quantum walks}, Nat. Phys. {\bf 13}, 1117 (2017).\\
26. X. Zhan, L. Xiao, Z. Bian, K. Wang, X. Qiu, B.C. Sanders, W. Yi, and P. Xue, {\it  Detecting topological invariants in nonunitary discrete-time quantum walks}, Phys. Rev. Lett. {\bf 119},
130501 (2017).\\
27. M. Pan, H. Zhao, P. Miao, S. Longhi, and L. Feng, {\it Photonic zero mode in a non-Hermitian photonic lattice}, Nat. Commun. {\bf 9}, 1308 (2018).\\ 
28. B. Midya, H. Zhao, and L. Feng, {\it Non-Hermitian photonics promises exceptional topology of light}, Nat. Commun. {\bf 9}, 2674 (2018).\\
29. S. Longhi, {\it Non-Hermitian topological phase transition in $\mathcal{PT}$-symmetric mode-locked lasers}, Opt. Lett. {\bf 44}, 1190 (2019).\\
30. S. Longhi. {\it Topological phase transition in non-Hermitian quasicrystals}, Phys. Rev. Lett. {\bf 122}, 237601 (2019).\\
31. B. Bahari, A. Ndao, F. Vallini, A. El Amili, Y. Fainman, and B. Kante, {\it Nonreciprocal lasing in topological cavities of arbitrary geometries}, Science {\bf 358}, 636 (2017).\\
32. M. A. Bandres, S. Wittek, G. Harari, M. Parto, J. Ren, M. Segev, D. N. Christodoulides, and M. Khajavikhan, {\it Topological insulator laser: Experiments}, Science {\bf 359}, eaar4005 (2018).\\
33. S. Longhi, {\it Non-Hermitian Gauged Topological Laser Arrays}, Ann. Phys. (Berlin) {\bf 530}, 1800023 (2018).\\
34. S. Mittal, E. A. Goldschmidt, and M. Hafezi, {\it A topological source of quantum light},  Nature {\bf 561}, 502 (2018).\\
35. H. Zhao, X. Qiao, T. Wu, B. Midya, S. Longhi, and L. Feng, {\it Non-Hermitian topological light steering}, Science {\bf 365}, 1163 (2019).\\
36. A. Regensburger, C. Bersch, M-A. Miri, G. Onishchukov, D. N. Christodoulides, and U. Peschel, {\it Parity-time synthetic photonic lattices}, Nature {\bf 488}, 167 (2012).\\
37. F. Cardano, A. D'Errico, A. Dauphin, M. Maffei, B. Piccirillo, C. de Lisio, G. De Filippis, V. Cataudella, E. Santamato, L. Marrucci, M. Lewenstein, and P. Massignan, {\it Detection of Zak phases and topological invariants in a chiral quantum walk of twisted photons}, Nat. Commun. {\bf 8}, 15516 (2017).\\
38. F. Dreisow, Y. V. Kartashov, M. Heinrich, V. A. Vysloukh, A. T\"unnermann, S. Nolte, L. Torner, S. Longhi, and A. Szameit, {\it Spatial light rectification in an optical waveguide lattice}, Europhys. Lett. {\bf 101}, 44002 (2013).\\
39. L. J. Maczewsky, J. M. Zeuner, S. Nolte, and A. Szameit, {\it Observation of photonic anomalous Floquet topological insulators}, Nat. Commun. {\bf 8}, 13756 (2017).\\
40. S. Mukherjee, A. Spracklen, M. Valiente, E. Andersson, P. Ohberg N. Goldman, and R. R. Thomson, {\it Experimental observation of anomalous topological edge modes in a slowly driven photonic lattice}, Nat. Commun. {\bf 8}, 13918 (2017).\\
41. A. Schreiber, K. N. Cassemiro, V. Potocek, A. Gabris, P.J. Mosley, E. Andersson, I. Jex, and Ch. Silberhorn,
{\it Photons Walking the Line: A Quantum Walk with Adjustable Coin Operations}, Phys. Rev. Lett. {\bf 104}, 050502 (2010).\\
42. A. D'Errico, F. Cardano, M. Maffei, A. Dauphin, R. Barboza, C. Esposito, B. Piccirillo, M. Lewenstein, P. Massignan, and L. Marrucci, {\it Two-dimensional topological quantum walks in the momentum space of structured light}, arXiv:1811.04001 (2019).\\
43. K. Mochizuki, D. Kim, and H. Obuse, {\it Explicit definition of $\mathcal{PT}$ symmetry for nonunitary quantum walks with gain and loss}, Phys. Rev. A {\bf 93}, 062116 (2016).\\
44. A. Guo, G.J. Salamo, D. Duchesne, R. Morandotti, M. Volatier-Ravat, V. Aimez, G. A. Siviloglou, and D. N. Christodoulides, {\it Observation of $\mathcal{PT}$-Symmetry Breaking in Complex Optical Potentials},
Phys. Rev. Lett. {\bf 103}, 093902 (2009).\\

\end{document}